\documentclass{sig-alternate}

\usepackage{times} %Added by Conrad because he prefers this font
\usepackage[numbers,sort&compress]{natbib} \usepackage{subfigure}
\usepackage{amssymb} \usepackage{amsmath} \usepackage{graphicx}
\usepackage[pdfpagelabels=false]{hyperref}
\usepackage{cleveref} %Use cleveref for references (i.e. use \cref instead of \ref)
\def\halfrs{\vspace*{-3ex}}
\def\rs{}
\newcommand{\xhdr}[1]{\paragraph*{\bf #1}}
\newcommand{\alg}{GCE}

\newcommand{\ie}{i.e.}

\newcommand{\Fscore}{F${}_1$-score }
\newcommand{\Fscores}{F${}_1$-scores }
\def\sharedaffiliation{%
\end{tabular}
\begin{tabular}{c}}

\begin{document}
\graphicspath{{pdffigs/}{epsfigs/}{mpsfigs/}}

% --- Author Metadata here ---
\conferenceinfo{The 4th SNA-KDD Workshop '10}{(SNA-KDD'10) July 25, 2010, Washington, DC USA}
\CopyrightYear{2010}
\crdata{978-1-4503-0225-8}
% \CopyrightYear{2007} % Allows default copyright year (200X) to be over-ridden - IF NEED BE.
% \crdata{0-12345-67-8/90/01} % Allows default copyright data (0-89791-88-6/97/05) to be over-ridden - IF NEED BE.
% --- End of Author Metadata ---

\title{Detecting Highly Overlapping Community Structure by Greedy
  Clique Expansion}

\numberofauthors{1} \author{ \alignauthor Conrad Lee, Fergal Reid,
  Aaron McDaid, Neil Hurley \sharedaffiliation
  \affaddr{University College Dublin}\\
  \affaddr{Clique Research Cluster}\\
  \affaddr{Dublin 4, Ireland}\\
  \email{\{conradlee,fergal.reid,aaronmcdaid\}@gmail.com,
    neil.hurley@ucd.ie} \date{30 January 2010} }
\maketitle
\begin{abstract}
  In complex networks it is common for each node to belong to several
  communities, implying a highly overlapping community
  structure. Recent advances in benchmarking indicate that the existing
  community assignment algorithms that are capable of detecting overlapping
  communities perform well only when the extent of community overlap
  is kept to modest levels. To overcome this limitation, we introduce
  a new community assignment algorithm called Greedy Clique Expansion
  (GCE). The algorithm identifies distinct cliques as seeds and
  expands these seeds by greedily optimizing a local fitness
  function. We perform extensive benchmarks on synthetic data to
  demonstrate that GCE's good performance is robust across diverse
  graph topologies. Significantly, GCE is the only algorithm to
  perform well on these synthetic graphs, in which every node belongs
  to multiple communities. Furthermore, when put to the task of
  identifying functional modules in protein interaction data, and
  college dorm
  assignments in Facebook friendship data, we find that GCE performs competitively.\\
\end{abstract}

\halfrs{} \vspace{5mm}
\noindent {\bf Categories and Subject Descriptors:}
H.2.8 {Database Management}: {Database Applications -- Data Mining}
% A category including the fourth, optional field follows...

\vspace{2mm}
\section{Introduction}
Community structure has been recognized in networks that come from a
wide range of domains, such as social and biological networks. While
concrete definitions of community vary by domain, a community may
generally be described as a set of nodes with dense internal
connections, exhibiting comparatively sparse connections to the rest
of the network.  Knowledge of community structure can reveal
functional organization in networks, much as identifying organs in the
body can reveal the role of various tissues.  In recent years,
numerous community assignment algorithms (CAAs) have been suggested,
as computer scientists and physicists have taken on the problem of
algorithmicly finding communities (for an excellent recent review of
the field, see \citet{fortunato-2010}).

Despite their proliferation, it is difficult to determine the
performance of CAAs for two reasons. On the one hand, there is a lack
of large empirical datasets where the \textit{a priori} or
\textit{ground truth} communities are known; and on the other hand,
most synthetic data---especially the most popular, the GN model
\cite{girvan-2002}---is overly simplistic and unrealistic, lacking key
topological features such as a heterogeneous degree distribution,
varied community sizes, and triadic closure, while also requiring that
every node belong to exactly one community. The lack of realistic
benchmark graphs has led to a situation where researchers know that
many algorithms perform well on simple networks, but are unaware how
these perform on more complex empirical data.

This problem is so pronounced that in his comprehensive review of the
field, \citeauthor{fortunato-2010} states with regard to benchmarking:
``...the issue of testing algorithms has received very little
attention in the literature on graph clustering. This is a serious
limit of the field. Because of that, it is still impossible to state
which method (or subset of methods) is the most reliable in
applications...''

In the last year, \citet{lancichinetti-2009:2} have addressed this
uncertainty by specifying a means of creating more realistic synthetic
benchmark graphs, which have scale-free degree and community size
distributions as well as overlapping communities. Using their
specification (called LFR), they and others have subsequently
discovered---with a level of subtlety previously unattained---under
what topological conditions a wide range of CAAs perform well or
poorly \cite{lancichinetti-2009:3,gregory-copra}.

\begin{figure}[t!h]
  \begin{center}
    \includegraphics[]{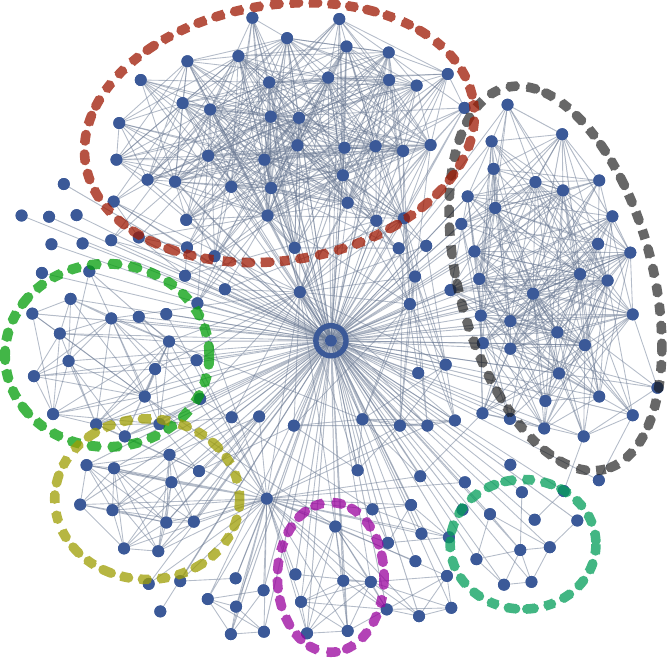}
    \caption{The ego-centric network of a Facebook user
      \cite{marlow-2009}. Note that this user belongs to several
      distinct communities, which we have outlined with dashed
      lines. }
    \label{facebook_egonet}
  \end{center}
  \vspace{-6mm}
\end{figure}

One surprising result revealed by this recent benchmarking is the poor
performance of many CAAs when it comes to detecting moderately
overlapping community structure.  It is intuitive from our knowledge
of real world domains that many complex networks will have communities
that overlap, potentially to a high degree.  Consider, for example, a
social network site like Facebook. On average, a Facebook user has 130
``friends,'' who typically belong to multiple distinct social groups
\cite{facebook-2010}. These groups may correspond to ties formed in
high-school, college, professional settings, and
family. \Cref{facebook_egonet}, which depicts the ego-centric network
of a Facebook user, demonstrates this tendency for a user to belong to
multiple groups.  The analysis of \citet{marlow-2009} suggests that
the groups apparent in this user's ego-centric network correspond to
acquaintances formed at different stages of life, and that most of
these groups are dormant. Clearly, if this type of ego-centric network
is typical of Facebook users, then any CAA that partitions nodes into
non-overlapping communities (henceforth, \textit{non-overlapping CAA})
will perform poorly: such CAAs can assign each node to only one of its
many communities.
Similarly, in complex networks of interactions between proteins, it
has been claimed that many proteins belong to multiple communities,
each of which in turn corresponds to some biological function
\cite{sawardecker-2009, palla-2005}. Since 2005, the year in which
\citet{palla-2005} published a groundbreaking CAA capable of detecting
overlapping communities, a number of algorithms have been developed
that are able to assign nodes to more than one community
\cite{palla-2005, clauset-2005,
  gregory-2007,gregory-2009:1,gregory-copra,
  mishral-2007,lancichinetti-2009, baumes-2005,shen-2009}. However,
using LFR networks \cite{lancichinetti-2009:3, gregory-copra} and
other synthetic networks \cite{sawardecker-2009}, recent work has
indicated that many CAAs that are supposed to be capable of detecting
overlapping community structure perform quite poorly when more than a
minority of nodes belong to multiple communities.

The purpose of this paper is two-fold.  On the one hand, in
\cref{method,complexity} we introduce a new algorithm, called Greedy
Clique Expansion (\alg)\footnote{\alg{} reference implementation
  available
  at:\url{http://sites.google.com/site/greedycliqueexpansion/}},that
is designed to perform well in domains with highly overlapping
community structure.  On the other hand, in
\cref{synthetic-benchmarks} we thoroughly benchmark this algorithm
alongside several other leading CAAs that are designed to detect
overlapping community structure.  We run benchmarks on graphs with
high levels of community overlap to clear up uncertainty of the
performance of CAAs designed for this domain.  None of these CAAs have
been subjected to such benchmarks. Our results indicate that GCE is
the only algorithm capable of accurately detecting communities when
nodes belong to several communities. We follow up these synthetic
benchmarks with two empirical benchmarks, one based on data from a
protein-protein interaction network, and the other on Caltech's
Facebook friendship network.

\section{Method}
\label{method}
Given a graph G with vertices V and edges E, \alg{} works by first
detecting a set of seeds in G, then expanding these seeds in series by
greedily maximizing a local community fitness function, and then
finally accepting only those communities that are not near-duplicates
of communities that have already been accepted.

In this section, we begin by reviewing the core concepts of \alg{}:
fitness functions, greedy expansion, seed selection, and community
distance measures. Finally, we fully specify \alg{}. Throughout, we
note how \alg{} differs from and resembles previous CAAs, and argue
why our modifications lead to better results on graphs that possess
highly-overlapping community structure, a claim which we substantiate
with benchmarking results in \cref{synthetic-benchmarks}.

\xhdr{Community fitness functions} A community fitness function $F:S
\rightarrow \mathbb{R}$ takes an induced subgraph $S$ of $G$ and
returns a real value that indicates how well $S$ corresponds to some
notion of community. In our discussion, the higher the value returned
by the fitness function, the better community
structure.  Many similar local fitness functions have been used in the context of
local CAAs
\cite{clauset-2005,baumes-2005,bagrow-2005,lancichinetti-2009,mishral-2007},
all of them different ways of formalizing the idea that a community's
internal edge density should be high when compared to its external
edge density.
\begin{figure}[t]
  \centering
  \includegraphics[scale=1.2]{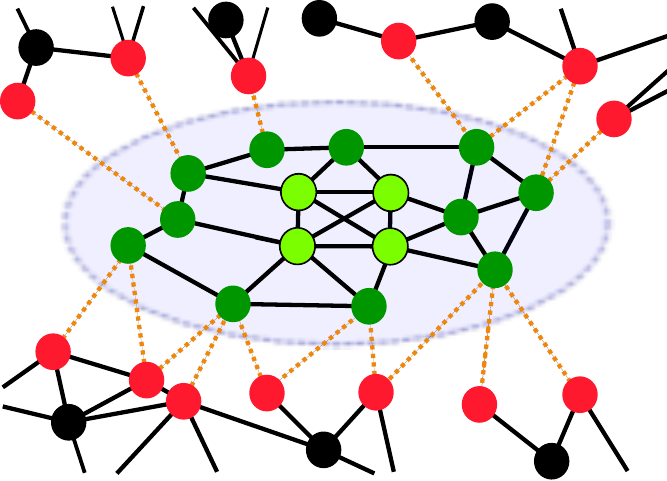}
  \caption{An idealized representation of a community, showing the
    seed clique in the center, surrounded by nodes added through
    greedy expansion.  The external edges, shown dashed, connect the
    community to its frontier.}
  \label{idealized_community}
  % \vspace{0mm}
\end{figure}

The fitness function is of central importance to the \alg{} algorithm;
it can be interpreted as steering the growth of a seed such that it
expands into the desired community. Just as there is no universally
correct concept of community that spans all domains, one cannot argue
that any given fitness function will be appropriate for all types of
network data.

Nevertheless, in our experiments, we found that the fitness function
defined by \citet{lancichinetti-2009} provided good results on a wide
range of synthetic and empirical data. They define the fitness of a
community $S$ in terms of $S$'s internal degree $k_{in}^{S}$ and
external degree $k_{out}^{S}$. $k_{in}^{S}$ is equal to twice the
number of edges that both start and end in $S$ (i.e., it is the sum of
the internal degrees of the nodes in $S$), and $k_{out}^{S}$ is the
number of edges that have only one end in $S$. In this notation,
they define community fitness as
\begin{equation}
  F_S = \dfrac{k_{in}^S}{(k_{in}^S + k_{out}^{S})^{\alpha}},
  \label{lancichinetti-fitness}
\end{equation}
where $\alpha$ is a parameter that can be tuned. Lower values of $\alpha$ result in larger
communities being fitter.  We found that $\alpha$ values in the range
$0.9-1.5$ provided the best results, which is in line with the
experience of \citeauthor{lancichinetti-2009} We resume discussion of
this parameter below.  \xhdr{Expanding a single seed} Assume, as
above, that $S$ is an induced subgraph of $G$ that can be thought of as
the ``seed'' or core of a community $C$. In other words, $S$ is
embedded in some larger community $C$, such that all of its nodes are
part of $C$, but not all nodes in $C$ are included in $S$.  The task
at hand is to expand $S$ by adding nodes to it until it includes all
nodes in $C$.  Previous work in community assignment suggests
that---by utilizing a community fitness function such as in
\cref{lancichinetti-fitness}--- $S$ can be efficiently expanded into
$C$ through a technique of greedy local optimization.
\cite{clauset-2005,baumes-2005,bagrow-2005,lancichinetti-2009} This
technique can be varied, but can be generally summarized in the
following steps:
\begin{enumerate}
\item For each node $v$ in the \textit{frontier} of $S$ (e.g., the red
  nodes in \cref{idealized_community}), calculate $v$'s node fitness,
  i.e., how much the addition of $v$ to $S$ would raise or lower the
  community fitness of $S$.
\item Select the node with the largest fitness, $v_{max}$.
\item If $v_{max}$'s fitness is positive, then add it to $S$ and loop
  back to step 1.  Otherwise, terminate and return $S$.
\end{enumerate}

The works cited above vary this technique by either using different
fitness functions or, for example, after each addition, removing any
nodes in $S$ if their removal would improve F(S). Also, they vary in
their approach of finding starting seeds. In general, this approach of
greedy local optimization scales well with the size and order of the
network because it works locally; we discuss complexity in more detail
in \cref{complexity}.

While \alg{} shares this general strategy of expanding seeds via
greedy local optimization, one key difference from previous algorithms
is the choice of starting seed.

\xhdr{Cliques as starting seeds} Various approaches have been used to
find the seeds of communities in the above-mentioned greedy
algorithms. \citet{lancichinetti-2009}'s LFM algorithm keeps randomly
selecting nodes that have not yet been assigned to any community,
until every node belongs to at least one community. This method
assumes that each node belongs to at least one community, and that, as
soon as every node has been assigned to at least one community, there
are no further communities that should be found. We take issue with
both of these assumptions; we believe that the latter---by implicitly
placing an upper bound on the number of communities that can be found
at $|V|$---is responsible for the declining performance of LFM as each
node tends to belong to more communities in the benchmarks that
follow.

\citet{baumes-2005}'s Iterative Scan algorithm selects random edges as
seeds, and keeps expanding seeds until the new seeds produce
communities that are duplicates of previously-found communities.
While this method of seed selection is less arbitrary than
\cite{lancichinetti-2009}'s, it is inefficient because, if one wants
to exhaustively search for seeds, the algorithm is unlikely to
terminate before it has expanded a vast number of duplicate
communities.

We use \textit{maximal cliques} (i.e., fully-connected sets of nodes
that are not completely contained in any larger set of fully-connected
nodes), as seeds; we will henceforth refer to maximal cliques simply
as cliques. This choice of seeds is motivated by the observation that,
on the one hand, cliques are one of the characteristic structures
contained within communities \cite{wasserman-1994}, while on the other
hand---if one discards smaller cliques that are highly embedded in
larger cliques--- they are rare structures. We note that other CAAs
exploit these properties of cliques \cite{palla-2005,yan-2009,
  shen-2009}, but none of them utilize cliques as seeds in the greedy
expansion strategy mentioned above.

One of the key parameters of our algorithm, $k$, is the minimum number
of nodes that a clique must contain if it is to be accepted as a
seed. On the one hand, $k$ should be sufficiently large such that any
clique of size $k$ or larger indicates the presence of a community;
otherwise, one risks expanding a seed into a region with no community
structure. For example, if $k=3$, then triangles can be accepted as
seeds. In some networks, one cannot assume that all triangles are
embedded in a community. In such cases, if $k=3$, then one risks
accepting seeds that are not embedded in any community, and these
seeds could expand into a region of the graph with no community
structure. Such communities can be thought of as false positives. On
the other hand, one should choose $k$ to be sufficiently small such
that all of the communities that one wishes to detect contain at least
one clique of size $k$. If $k$ is chosen to be too large, then those
communities that ought to be detected, but contain no sufficiently
large cliques, will not be detected. Such cases could be considered
false negatives. We find that $k$ values of 3 or 4 will generally
satisfy these constraints, and that one can decide between these two
possibilities by considering whether one's preferred semantic
definition of community includes small structures.

This choice of seeds comes with an implicit requirement that any
community that can be found by \alg{} must contain a clique with $k$
or more nodes.  One might object that this assumption is too strict
and will result in many false negatives. However, our results on the
LFR benchmark graphs indicate otherwise.  The LFR benchmarks provide a
challenge in this respect because the process which creates
communities does not favor the generation of cliques. In particular,
creates edges such that
the probability that any pair of nodes in a community is connected by
an edge is independent of whether those two nodes share neighbors,
much as in a classic Erd\H{o}s-R\'{e}nyi random graph
\cite{fortunato-2010}.  This generation technique does not lead to the
high number of triangles and cliques that have been observed in
empirical graphs.  Empirical networks show a strong tendency for
transitivity, i.e., for two neighbors of a given node to be connected
to each other \cite{watts-1998,serrano-2006}, a process which leads to
higher clustering coefficients and more cliques than one would expect
to find in correspondingly sparse Erd\H{o}s-R\'{e}nyi graphs. Thus,
the fact that \alg{} performs well on these synthetic graphs---despite
the fact that one expects fewer cliques in these than in empirical
data---indicates that this minimum clique size requirement does not
cripple the sensitivity of \alg.

\xhdr{Duplicates and community distance} Overlapping CAAs must include
some implicit or explicit strategy for dealing with near-duplicate
communities (in contrast to the more numerous, non-overlapping CAAs,
whose communities can share no nodes). This problem arises from the
fact that many seeds can grow into near-duplicate versions of the same
community. This is undesirable from the perspective of the network
analyst because results that contain a large number of near-duplicate
communities are hard to interpret and report statistics on.

To rid our results of near-duplicate communities, we must formally
define what we mean by near-duplicate communities. Along the lines of
\citet{baumes-2005}, we begin by defining a community distance
measure. We choose a symmetric measure of community distance that can
be thought of as the \textit{percent non-embedded}. Given two
communities $S$ and $S'$, this measure is defined as
\begin{equation}
  \delta_E(S,S')= 1 - \dfrac{|S \cap S'|}{ \min (|S|,|S'|)},
  \label{percent-non-embedded}
\end{equation}
which can be interpreted as the proportion of the smaller community's
nodes that are not embedded in the larger community.

Given a set of communities $W$ and a community $S$, we can define the
near-duplicates of $S$ to be all communities in $W$ that are within a
distance $\epsilon$ of $S$, where $\epsilon$ is the minimum community
distance parameter.

\vspace{-1mm} \xhdr{Overview of \alg} Now that we have covered the
requisite concepts of community fitness, expanding a seed, choosing
seeds, and near-duplicate seeds, we can outline the \alg{}
algorithm. Given a graph $G$, a minimum clique size $k$, a minimum
community distance $\epsilon$, and a scaling parameter $\alpha$, our
algorithm:

\begin{enumerate}
\item Finds seeds by detecting all maximal cliques in $G$ with at
  least $k$ nodes.
\item Creates a candidate community $C'$ by choosing the largest
  unexpanded seed and greedily expanding it with a community fitness
  function $F$ until the addition of any node would lower fitness.
\item If $C'$ is within $\epsilon$ of any already accepted community
  $C$, then $C$ and $C'$ are near-duplicates, so discard
  $C'$. Otherwise, if no near-duplicates are found, accept $C'$.
\item Continues to loop back to step 2 until no seeds remain.
\end{enumerate}

We note that although \alg{} allows the user to specify the values of
three parameters, two of these---$k$ and $\epsilon$---allow for
versatile default values.  The value of $k$ should usually be $4$; if
one is interested in very small communities (as in the case of the
protein complexes presented in \cref{empirical-benchmarks}), then $k$
should be set to $3$. We find $0.25$ to be a good default value for
$\epsilon$---if one finds too many near-duplicate communities in the
output, then $\epsilon$ should be increased.

The scaling parameter $\alpha$ lends itself least to a versatile
default value. For best results, one should first run \alg{} with
$\alpha$ set to $1.0$, look at the results, and decide whether the
communities found by \alg{} should be larger (using a lower $\alpha$)
or smaller. However, in cases where the user knows little about the
community structure, such tuning may be difficult. For this reason, in
the synthetic and empirical benchmarks that follow, we report results
where $\alpha$ is set to $1.0$, rather than tuning this parameter for
best results.  The good benchmarking results indicate that even if one
cannot tune $\alpha$, GCE will often return good results with all
parameters set to their default values.

\section{Optimizations}
\label{complexity}

We begin this section by underscoring the point that one cannot
satisfyingly characterize the average complexity of \alg{} purely in
terms of $|V|$ or $|E|$; rather, the complexity depends on subtler
local characteristics of $G$ that are difficult to specify rigorously. Despite
this, we can clearly discuss several important parts of the algorithm,
and the heuristics and optimizations we have used to improve their
performance.

First, we discuss finding the maximal cliques that form seeds, and
consider the complexity of greedily expanding each seed. The remainder
of this section is devoted to various problems related to detecting
near-duplicate communities: We
consider the cost of deciding whether a candidate community is
sufficiently distinct to accept, and two heuristics to discard
potentially indistinct seeds early. While our solutions to these
problems are somewhat trivial, we describe them in detail both for
replicability and because the computational savings that they afford
are so significant. 

We note that \alg{} makes
extensive use of set operations, such as set insertion and deletion
and testing for set membership. In our implementation, we use the C++
standard template library to provide these operations.

\xhdr{Finding Maximal Cliques} Although finding all of the cliques in
a graph is generally computationally expensive, cliques can be found
quickly in graphs that are sufficiently sparse. To this end, our
implementation makes use of the Bron-Kerbosch \cite{BronKerbosch}
clique enumeration algorithm to efficiently find the maximal cliques that
form seeds.  In the large synthetic and empirical networks that we
analyze in \cref{synthetic-benchmarks} and
\cref{empirical-benchmarks}, the computation required for finding
cliques was a small part of the overall run time, compared to the
computation required to expand seeds and check for near duplicates.
To futher support the claim that finding cliques in sparse graphs is
scalable, we point out that \citet{Schmidt2009} have recently introduced a parallel
variant of the Bron-Kerbosch algorithm, which they demonstrate can achieve a linear parallel
speed-up even when using 2048 processors.

\xhdr{Greedy Expansion} Greedy seed expansion requires that the
frontier $f(S)$ of each initial seed be identified, and that $f(S)$ be
updated as $S$ is expanded.

The initial frontiers may be identified by calculating, for each edge,
the symmetric difference of the sets of seeds for the endpoints of
that edge.  Identifying the initial frontiers therefore has complexity
$O(|E|\times M)$, where $M$ is the number of cliques to be expanded.

As each seed is expanded by adding the fittest node from the frontier $v_{max}$, its frontier becomes:
\[
f(S \cup v_{max}) = (f(S) \cup N(v_{max})\backslash S) - \{v_{max}\}
\]
where $N(v_{max})$ is the set of neighbors of $v_{max}$.  This
requires at most $\theta$ insertions into $f(S)$, where $\theta$ is
the maximum degree in $G$.  We note that the fitness $F_{S \cup
  \{v_{max}\}}$ depends only on the internal and external degree of
$v_{max}$ and the total internal and external degrees of the vertices
already in S.  To facilitate fast identification of $v_{max}$,
$k_{in}^S$ and $k_{out}^S$ are stored, along with the internal and
external degrees of each $v$ in $f(S)$.
% $k_{in}^v$, $k_{out}^v$ for
These stored values are updated each time $f(S)$ is updated, noting
that $k_{in}^S$ and $k_{out}^S$ each change only by the internal and
external degree of $v_{max}$ and that the frontier degrees need only
be updated for $v \in N(v_{max})$.

\xhdr{Identifying Near-duplicate Communities} To identify if a
candidate community, $C$, is a near-duplicate of an accepted
community, it is necessary to calculate the overlap between community
pairs.  Finding the intersection between sets $C_1$ and $C_2$ has
complexity $2(|C_1|+|C_2|)-1$, assuming sorted sets.  In a naive
implementation, after each seed expansion has completed, this must be
carried out $O(\zeta)$ times, with $\zeta$ the number of
accepted communities, resulting in at least $O(\zeta^2)$ set intersetions
 in total.  Instead, we maintain, for each node $v$, the
set $c(v)$ of accepted communities it belongs to.  In a first pass, we
identify those communities that have a non-empty overlap with the
candidate community as $ \bigcup_{v \in C} c(v)$.  The full
intersection is then calculated only on the communities with non-empty
overlap.  At the cost of extra storage, this results in significant
time savings as the number of accepted communities grows large.

\xhdr{Clique Coverage Heuristic (CCH)} While developing this
algorithm, we observed that many empirical datasets exhibited the
following phenomenon: given some maximal clique, there exist numerous
smaller cliques that are almost, but not fully, subgraphs of
it. \citet{shen-2009} have also observed this property; in fact, this is the key property exploited by
k-clique percolation. \cite{palla-2005} (Remember, whenever we use the term ``clique'' here and
throughout our paper, we mean ``maximal clique.'')  Typically, these
cliques are within $\epsilon$ distance of the larger clique and thus
are near-duplicates of it.

We also observed that when expanded, such sets of near-duplicate
cliques were likely to grow into the same or very similar regions of
the graph, leaving us with many near-duplicate communities that would
later need to be removed (as specified in subsection
``\textit{Duplicates and community distance}'' above). We could thus
skip the expansion of these near duplicate seeds without significantly
affecting our results and enjoy a significant savings in memory and
computation time.  We developed a simple heuristic to quickly prune
such near-duplicate cliques from the our initial set of seeds.  This
method is designed solely as a heuristic computational speedup, and
has been found to effectively discard large numbers of near-duplicate
cliques while not significantly altering benchmark results.
\begin{figure}[t]
  \centering
  \includegraphics[scale=1.0]{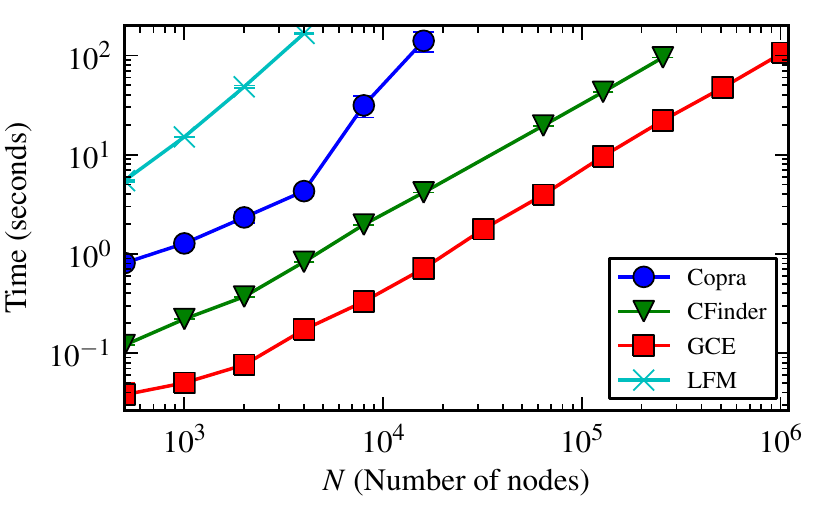}
  \caption{With constant degree and community size, \alg{} operates on
    over ten million edges in two minutes.}
  \label{increasing_nodes}
  \vspace{-2mm}
\end{figure}
\begin{figure}[]
  \centering
  \includegraphics[scale=1.0]{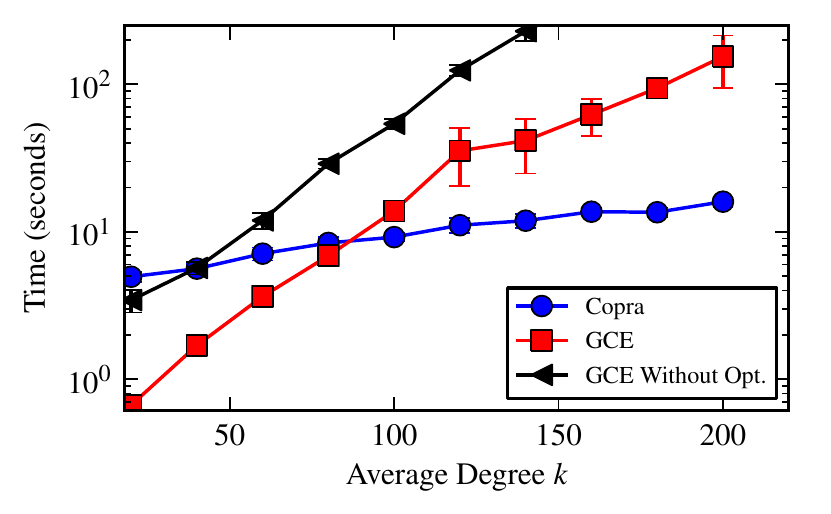}
  \caption{In even relatively small graphs, which contain $5000$
    nodes, \alg's performance degrades as degree increases. The CCH
    optimization increases \alg's scalability.}
  \label{increasing_degree}
  \vspace{-2mm}
\end{figure}

The Clique Coverage Heuristic (CCH) is as follows: We order the maximal
cliques, largest first. Each clique is then either accepted or
rejected.  A clique is accepted if less than a certain proportion
$\phi$ of its nodes are contained in at least two of the larger cliques that
have already been accepted.  We found that even values of $\phi$ close
to 0 resulted in the large numbers of near-duplicate cliques being
rejected from consideration.  Simultaneously, it appears that in each
of the clique-dense areas of the graph that are likely to be embedded
in a community, at least some of the original cliques are always
preserved. Thus, each of these areas remains ``seeded'' with at least
one clique, but most of the smaller, near-duplicate versions of this
clique are removed. In our evaluation we choose a value of .25 for
$\phi$ -- meaning that a clique will only be discarded if at least 75
percent of its nodes have already been covered twice by other, larger,
accepted cliques.

Due to the potentially vast number of near-duplicate cliques found in
complex networks, the impact of CCH can be large. For example, we ran GCE on
the Oklahoma State Facebook subnetwork, which contains 17425 nodes and
829528 edges (the source of this data set is
\citet{traud-2009}, which is described in
\cref{empirical-benchmarks}). On this data, CCH reduced the number of
cliques from over 46 million to around 5000. To address concerns that pruning seeds using CCH may adversely affect
the accuracy of GCE, we refer the reader to \cref{synthetic_overlapping}, which includes
one line displaying the results of GCE with the CCH, and one line for
GCE without the CCH. The results are almost identical.

\xhdr{Abandoning suspiciously overlapping seeds} Let us call any seed
that is undergoing expansion that is within some distance $\Delta$ of
an already accepted expanded community `suspicious'. We call it
suspicious because for an appropriate value of $\Delta$, such
expanding seeds frequently expand to within $\epsilon$ (i.e. become
near-duplicates) of an already accepted community.  In these cases,
the computation required to expand the suspicious seed into a
near-duplicate community is essentially wasted. One simple
optimization, which we employ in all of the evaluations below, is to
discard these suspicious seeds.  In practice, we have found that the
benchmark results of our implementation are not very sensitive to a
range of $\Delta$ values, and we use the value of .6 for both $\Delta$
and $\epsilon$.

\xhdr{Performance characteristics}
We now briefly describe two experiments to
reveal both a strength and a weakness of \alg's runtime performance
(we postpone discussion of accuracy, as this is the topic of the next section). These experiemnts are
based on the above-mentioned LFR graphs, which are synthetic graphs whose
specification will be explained in detail in
\cref{synthetic-benchmarks}. The parameters used to constuct the
graphs used in these experiments are listed in the first two columns of
\cref{parameters}.

While the details of these graphs may not be
clear for the reader until the next section, the essential point is
that in \cref{increasing_nodes}, we run \alg{} on a series of graphs
that are identical in many ways---sharing the same degree distribution
and community size distribution---but which have ever-more nodes and
communities. In this figure we observe that as the size of the graph
increases along the x-axis, the runtime of \alg{} scales
favorably when compared to the runtime of other overlapping CAAs
(which are also introduced in the next section).

In \cref{increasing_degree} we again observe the runtime of \alg{} on a
series of graphs that are nearly identical; however, in these graphs
we do not vary the number of nodes (keeping it fixed at 5000), but rather we vary the degrees of
the nodes. As the average degree increases along the x-axis, the
runtime of \alg{} increases quite rapidly when compared with the
quickest algorithm run on this experiment, COPRA (which is also
introduced in the next section). We observe
that even on these relatively small graphs, the runtime of \alg{}
is quite sensitive to the average degree. In this figure, we also see
the effect of the CCH optimization.

\section{Synthetic Benchmarks}
\label{synthetic-benchmarks}
In the following section, we pursue two objectives.  First, we
perform the central experiment of this paper: we benchmark many CAAs
on the synthetic graphs in which nodes belong to to up to five communities.
Benchmarks with such a high degree of community overlap are uncharted
territory.  The results of the benchmarks are surprising, indicating that many
algorithms that are specifically designed for detecting overlapping
community structure perform poorly when more than a fraction of nodes
belong to more than one community. While one other algorithm performs
well up to the point where nodes belong to an average of $1.9$
communities, \alg{} returns good results even when every node belongs
to four communities.

Unfortunately, because benchmarks with such high levels of communities
have not yet been carried out, we had to
choose many of the parameters for creating the benchmark
graphs. The secondary purpose of this section is to show that \alg{} also
performs well on benchmarks graphs whose parameters have been defined
by other, less interested parties.  To this end, we replicate a set of partitioning benchmarks
performed by \cite{lancichinetti-2009:3} in a recent review of several
CAAs. We demonstrate that \alg{} performs competitively against the
best known CAAs. This is significant, because these other algorithms
are specialized for graph partitioning, while \alg{} is designed to
handle the more general case of overlapping communities.

\xhdr{Benchmarking procedure and terminology}To benchmark the
performance of a CAA, we perform the following steps: first, we create
a synthetic graph which, by construction, contains communities planted
in it. We will refer to these communities as the \textit{ground truth}
communities. Next, we run a CAA on this graph; we call the communities
returned by the CAA the \textit{found} communities. Finally, we use
some metric to compare the similarity of the ground truth communities
to the found communities.

To construct synthetic graphs, we use the LFR specification, which allows
one to create graphs with realistic properties such as scale-free degree and
community size distributions. \citep{lancichinetti-2009:2}. To measure the similarity of ground
truth communities and found communities, we use normalized mutual information
(NMI), an information-theoretic similarity measure.  This measure is normalized such that
the NMI of two sets of communities is 1 if they are identical, and 0
if they are totally independent of each other. \citet{danon-2005}
first applied NMI
to the problem of evaluating the similarity of two sets of
communities, but defined the measure only for partitions. In our
benchmarks, we employ a variant of NMI introduced by
\citet{lancichinetti-2009} that is defined for covers, in which nodes may belong to multiple communities.\footnote{For creating the
  LFR graphs and measuring overlapping NMI, we use the implementations
  provided by the authors, both of which are freely available at
  \url{http://sites.google.com/site/andrealancichinetti/software}.}
\begin{figure*}[t]
  \subfigure[]{
    \includegraphics[]{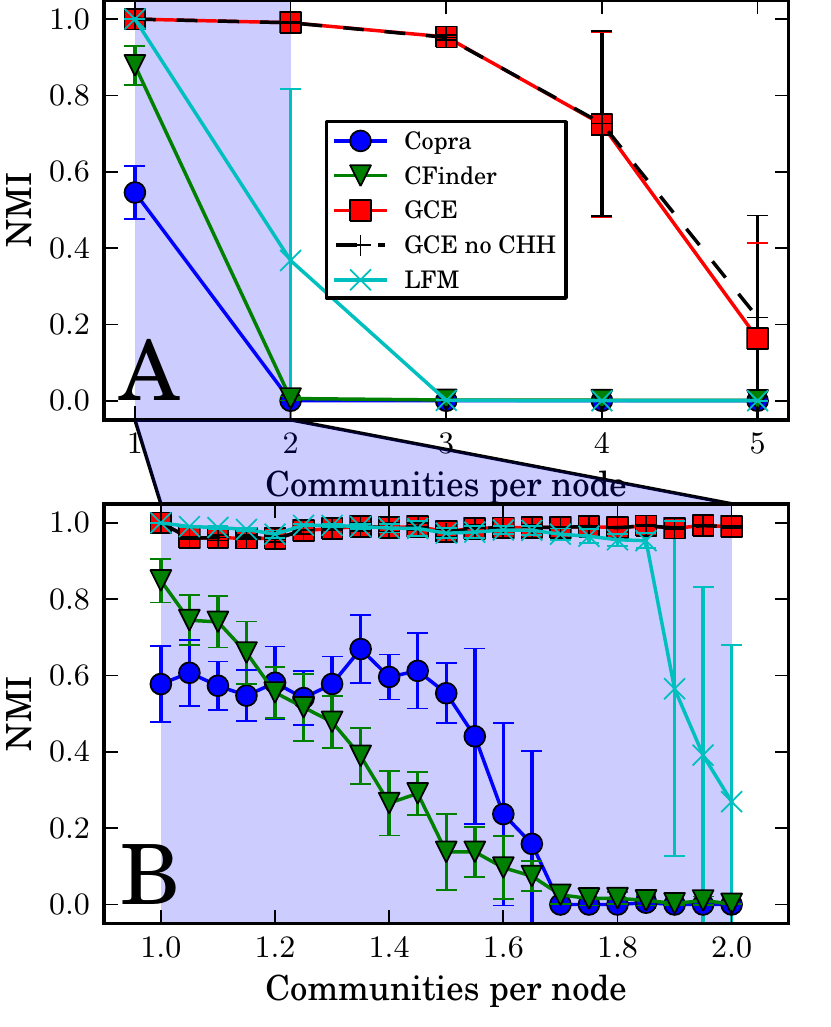}
    \label{synthetic_overlapping}
  }\hspace{0.0in} \subfigure[]{
    \includegraphics[]{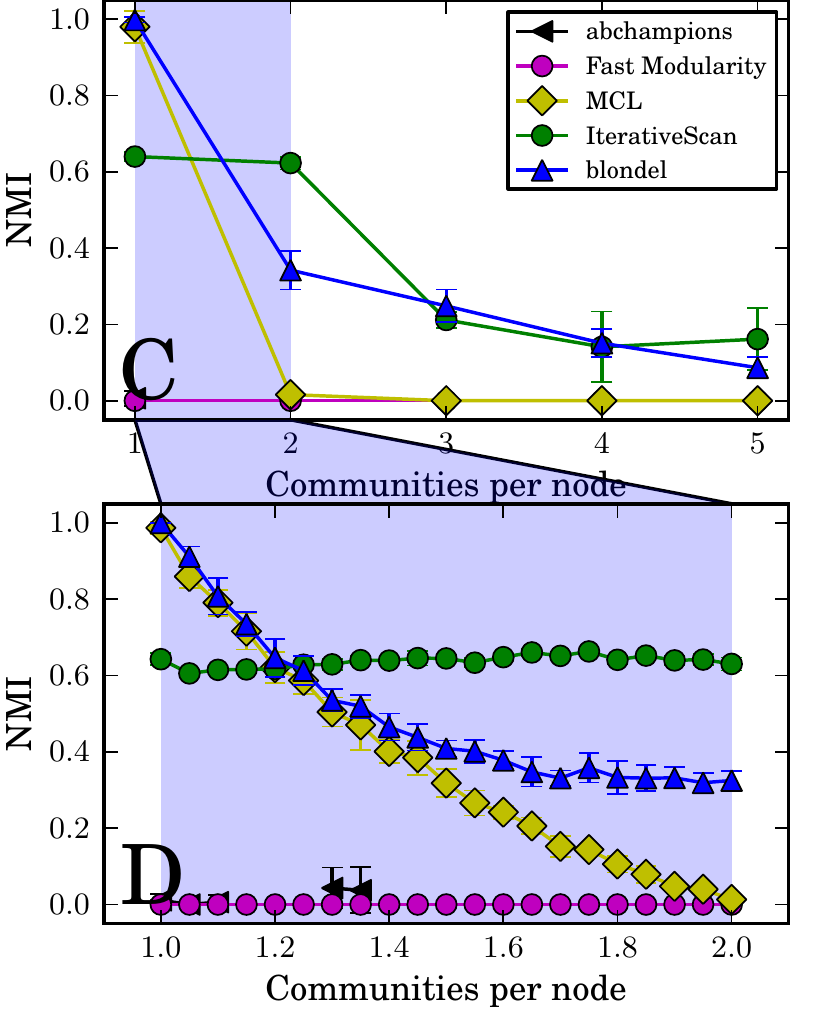}
    \label{overlapping_additional}
  }
  % \vspace{-5mm}
  \caption{Accuracy of nine CAAs on LFR graphs with increasing amounts
    of community overlap.}
  % \vspace{-5mm}
\end{figure*}

\xhdr{Synthetic graph description \& parameters} To construct a LFR
graph, one must specify ten parameters, which are listed in
\cref{parameters}.  Note that the parameter $O_{n}$ refers to the
number of nodes in the graph that are overlapping (i.e., belong to
more than one community), and $O_{m}$ dictates how many communities
each of the overlapping nodes belongs to.

\begin{table}[]
  \centering
  \setlength{\tabcolsep}{1pt}
  \begin{small}	
    \begin{tabular}{|ll|cccc|}

      \hline
      \textbf{}	& \textbf{Description}  & \textbf{\cref{increasing_nodes}} & \textbf{\cref{increasing_degree}} & \textbf{\cref{synthetic_overlapping}} & \textbf{\cref{synthetic_disjoint}}
      \\ \hline    \hline
      $N$ & number of nodes                  		& \textbf{500-1mil} 	& 5000 			& 2000 		& 1000/5000
      \\\hline $k$          & average degree 		& 10			& \textbf{20-200} 	& \textbf{18-90}& 20 
      \\\hline $k_{max}$    & max degree 		& 40 			& $200$			& 120 		& 50 
      \\\hline $C_{min}$    & min comm. size 		& 10 			& $k$			& 60 		& 10/20 
      \\\hline $C_{max}$    & max comm. size 		& 50			& $500$			& 100		& 50/100 
      \\\hline $\tau_{1}$   & degree exponent 	& $-2$ 			& $-2$			& $-2$ 		& $-2$ 
      \\\hline $\tau_{2}$   & comm. exponent 		& $-1$ 			& $-1$ 			& $-2$ 		& $-1$ 
      \\\hline $\mu$        & mixing parameter 	& 0.4			& $0.4$ 		& .2 		& \textbf{0.1-0.85} 
      \\\hline $O_{n}$      & num. overlap nodes 	& $k/2$			& $0$ 			& 2000 		& 0 
      \\\hline $O_{m}$      & comms per node 		& 2 			& $1$			& \textbf{1-5} 	& 1 
      \\ \hline   

    \end{tabular}
  \end{small}
  \caption{Parameters for LFR synthetic benchmark graphs for various
    figures. Bolded values indicate the the variable 
    on the x-axis of each plot.}
  \label{parameters}
  \vspace{-4mm}
\end{table}

There are two types of edges in LFR graphs: those that are planted
within communities, and those that are created randomly independent of
community structure.  The mixing parameter $\mu$ controls the
proportion of random edges to total edges; for example, if $\mu=0.2$,
then the LFR algorithm creates a graph such that approx. 80\% of each
node's edges end within that node's communities, and the remaining
20\% end in some randomly selected community. In general, as $\mu$
increases to 1, the community structure becomes ever weaker.

\xhdr{Benchmarks on graphs with overlapping communities} As we
mentioned in the introduction, very little work has been done on
benchmarking CAAs on graphs with overlapping community structure. The
work that includes such benchmarks is limited to graphs with only
moderate levels of community overlap, \ie, where some fraction of
nodes belong to two communities, and the rest belong to one
\cite{lancichinetti-2009:3, gregory-copra,sawardecker-2009}.

Our purpose here is to examine how CAAs perform on graphs with a
higher degree of community overlap. We specify five graphs; in the
first, each node belongs to one community, and in each successive
graph all nodes belong to one more community, so that in the fifth
graph, all nodes belong to five communities.  In order for nodes to
have enough edges for membership in an increasing number of
communities, in each successive graph nodes are assigned a higher
average degree.  In the first graph, the average node degree is
$k=18$, and for each extra community that a node belongs to, the
average degree increases by $18$, so that by the fifth graph,
$k=90$. Other parameter values remain constant across all five graphs
and are listed in the third column of \cref{parameters}. In all
synthetic and empirical benchmarks, we kept \alg's parameters fixed at
$k=4$, $\alpha=1.0$, and $\epsilon=0.6$, with the exception that in
the PPI benchmark, we set $k=3$ because the domain contains very small
communities.
 
\begin{figure*}[]
  \vspace{-6mm} \centering \subfigure[Large $C$, Small $G$]{
    \includegraphics[]{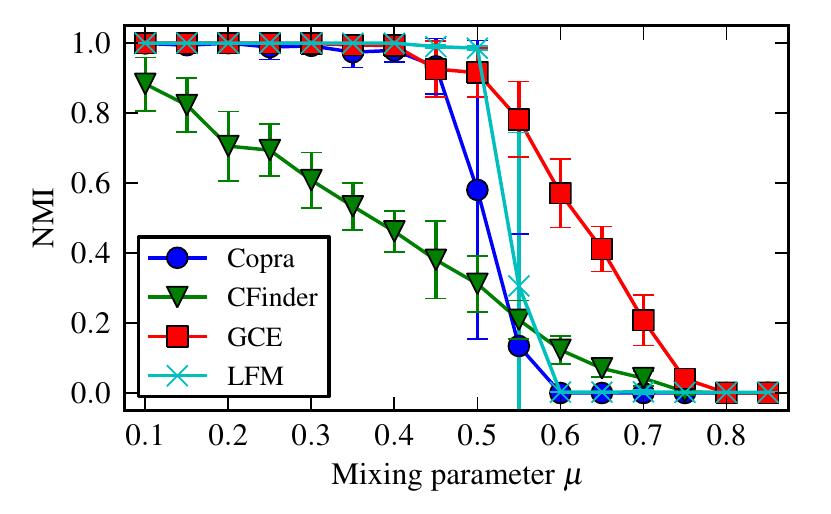}
    \label{appendixsynthetic}
    \label{synthetic_disjoint_large_small}
  } \subfigure[Large $C$, Large $G$]{
    \includegraphics[]{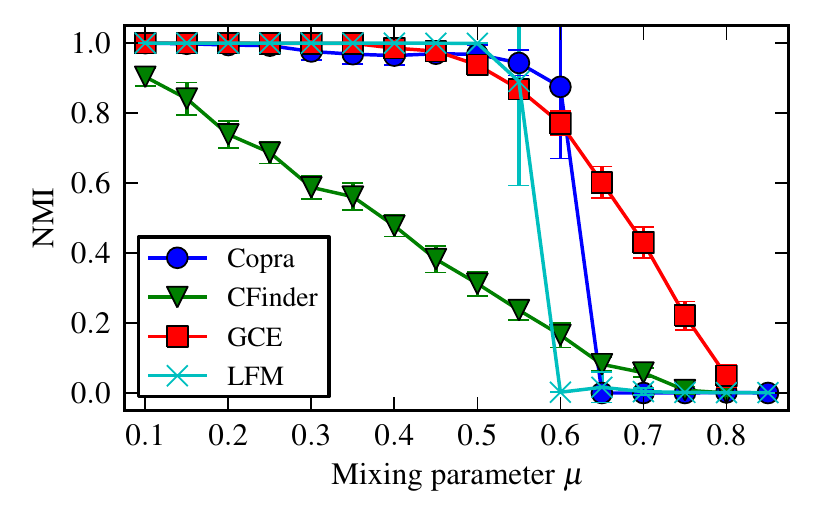}
    \label{synthetic_disjoint_large_large}
  } \subfigure[Small $C$, Small $G$]{
    \includegraphics[]{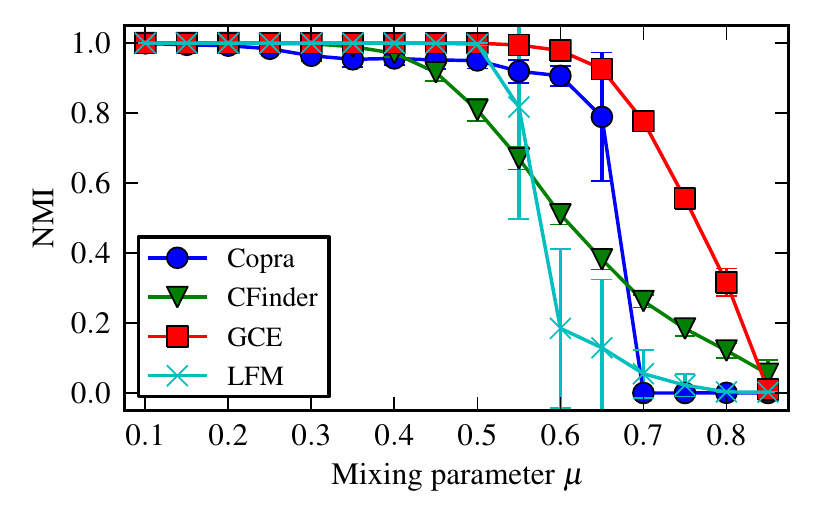}
    \label{synthetic_disjoint_small_small_2}
  } \subfigure[Small $C$, Large $G$]{
    \includegraphics[]{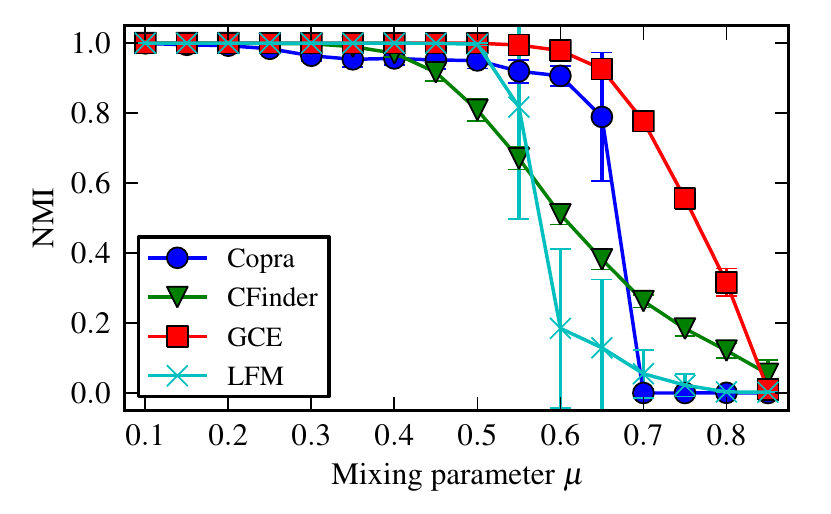}
    \label{synthetic_disjoint_small_large}
  }
  \caption{\label{synthetic_disjoint} NMI of \alg{} and other
    overlapping CAAs on all benchmarks used by
    \citet{lancichinetti-2009:3} and \citet{gregory-copra}.}
  \vspace{-4mm}
\end{figure*}

In \cref{synthetic_overlapping} (A) and (B), we see the results of
\alg{} alongside five other CAAs that were designed specifically to
detect overlapping communities, and two that find non-overlapping
communities. The five other overlapping CAAs that we benchmarked are
CFinder, which employs a technique of k-clique percolation
\cite{palla-2005}; LFM, which uses a local greedy optimization
strategy very similar to GCE, but selects seeds randomly
\cite{lancichinetti-2009}; COPRA, which utilizes a label propagation
technique \cite{gregory-copra}; abchampions, which finds all regions
of the graph with a certain difference between internal density and
external sparsity \cite{mishral-2007}; and Iterative Scan
\cite{baumes-2005}, which we have described in \cref{method}. All
implementations we used were from the authors. Just as \alg's
parameters were fixed at default values, we left the parameters of the
other algorithms set to their defaults with the following exceptions:
CFinder, where we set $k=4$, which returned the best results overall;
COPRA, where we set $v=3$, Iterative Scan, where we set the initial
cluster size to 2, as recommended; and abchampions, where we set
$C_{min}=5$ and $C_{max}=100$, as recommended by an author.  For each
point in the plot, we ran all CAAs on the same 10 realizations of
graphs created with the same parameters; error bars represent the
standard deviation of NMI over the 20 runs, which also holds for the
benchmarks in \cref{synthetic_disjoint}.

\Cref{synthetic_overlapping} (A) and (C) suggests why no benchmarks
have been carried out on graphs with such a high degree of overlap:
none of the existing algorithms perform well even when overlap is kept
at moderate levels. The robust performance of \alg{} as the number of
communities to which each node belongs increases is 
unprecedented. To more closely examine the limitations of the other
three algorithms, in \cref{synthetic_overlapping} (B) and (D) we run
additional benchmarks on graphs where some fraction of nodes belongs
to two communities and the rest belong to one. In these
graphs, the average degree $k$ steadily rises from $18$ (when each
node belongs to one community) up to $36$ (when every node belongs to
two communities).

With regard to CFinder and COPRA, our results here mirror the results
of \cite{lancichinetti-2009:3,gregory-copra}, who benchmarked these
algorithms on similar LFR-generated graphs. They also found that these
algorithms could no longer accurately assign nodes to communities if
all nodes belonged to two communities.  We benchmark LFM for the first
time on such overlapping synthetic graphs. It is interesting to note
that although LFM and \alg{} use the same fitness function and a
similar greedy heuristic, their
results vary so greatly.
uses the same fitness function as \alg{}. We speculate that LFM's performance
drops because this seed selection strategy causes it to prematurely
give up on trying to expand new regions of the graph that have
unidentified communities.

\xhdr{Benchmarks on graphs with disjoint communities} Most of the
benchmarking of CAAs has been on graphs with non-overlapping
communities. In particular, \citet{lancichinetti-2009:3} have
benchmarked a wide variety of CAAs on a particular set of LFR
graphs. \citet{gregory-copra} have recently followed suit and
benchmarked more algorithms on this set of graphs, so we continue in
this vein and benchmark \alg{} and a number of other CAAs to see how
they perform on this emerging standard.  There are four graph
specifications that are included in this standard set.  The graphs in
this quartet are either small or large ($N=1000$ or $N=5000$), and
have either small or large communities (ranging between $10-50$ nodes
or $20-100$ nodes).  The results are displayed in \cref{synthetic_disjoint}.
% \begin{figure}
%   \includegraphics[]{disjoint_small_comm_small_graph_10iter}
%   \caption{Results of four CAAs on a LFR benchmark specification where
%     $N=1000$ and community sizes are $10-50$ nodes.}
%   \label{synthetic_disjoint_small_small}
%   \vspace{-4mm}
% \end{figure}

It is informative to compare the accuracy of \alg{} as displayed in
\cref{synthetic_disjoint} 
with the results of various disjoint CAAs in the recent comparative
benchmarking of \citet{lancichinetti-2009:3}. We note that COPRA and
CFinder have been previously benchmarked on graphs with the same
specification and returned the same results, indicating that we have
accurately replicated this benchmark and that it is reasonable to
compare our results with theirs.  With regard to this matter of
comparison, we also point out that in their review,
\citeauthor{lancichinetti-2009:3} also used the overlapping version of
NMI, so that we can directly compare the results from
\cref{synthetic_disjoint} to their results.  The
comparison indicates that \alg's accuracy on the task of partitioning
graphs with non-overlapping community structure is among the best,
even when compared to non-overlapping CAAs, which specialize in this
task.

More specifically, \alg{} clearly outperforms the classic divisive GN
algorithm of \citet{newman-2004-2}, a similar divisive algorithm by
\citet{radicchi-2004}, the EM method of \citet{newman-2007}, the
Markov clustering algorithm (MCL) of \citet{vandongen-2000:2}, an
information theoretic approach by \citet{rosvall-2008}, and a spectral
algorithm by \citet{donetti-2004}.  Against other algorithms, results
were mixed.  \alg{} performed better than a method based on modularity
optimization via simulated annealing by \citet{guimera-2005} in all
cases except where the graph size was small and the community size
large. \alg{} performs similarly to the modularity maximizing
algorithm of \citet{blondel-2008}, which was among the CAAs that
\citeauthor{lancichinetti-2009:3} identified as a top performer, and
slightly worse than the other two top performers: another information
theoretic algorithm from \citet{rosvall-2008}, and a Potts model
approach by \citet{ronhovde-2009}.

\section{Empirical Benchmarks}
\label{empirical-benchmarks}
In this section we strive to demonstrate GCE's ability to identify
meaningful communities in the context of non-trivial empirical networks, for which
ground-truths are available. We agree with \cite{fortunato-2010} that
small the social networks which are typically
used as empirical benchmarks for CAAs, such as Zachary's Karate club, provide insufficient validation of a
CAA. While we were unable to find an ideal, large-scale graph in which
the ground-truth is completely known, we find two reasonable data sets
for this purpose: a protein-protein
interaction (PPI) network which includes a set of known protein
complexes, and a collegiate Facebook network in which the dorm
assigmnets are known.
\xhdr{Protein-protein interactions and protein complexes} Protein
complexes tend to correspond to groups of proteins with many
interactions, and can thus be detected by CAAs. We use a set of known
protein complexes as an approximate, ground truth.

To construct the PPI network, we used the interaction data found in
the Combined-AP/MS network \cite{CombinedAPMS}.\footnote{ Available at
  \url{http://interactome.dfci.harvard.edu/S_cerevisiae/}}, which
contains $1622$ proteins and $9070$ interactions.  For the ground
truth communities, we used the complexes listed in the CYC dataset of
known complexes\footnote{ Available at
  \url{http://wodaklab.org/cyc2008/}}, selecting only those complexes
that were also in the PPI network.  Because many of these complexes
were simply edges or triangles that are not recognizable as network
communities, we removed all complexes with fewer than four proteins
from the ground truth. Consequently, we use the value of 3 as the
minimum clique size for all clique based algorithms. Note also, we use
SCP\cite{SCP} instead of CFinder here, as the latter fails to
terminate on this dataset.  The resulting ground truth contains $880$
proteins; $136$ of these belong to more than one complex.
As the ground truth contains only a modest amount of overlap, we
compare GCE against some disjoint CAA algorithms as well as some
overlapping CAA algorithms.  In \cref{ppi-nmi}, we show the NMI
achieved by each algorithm, along with the number of communities
found.

% What's the third column??
We see that \alg's found communities have the highest NMI with the
ground truth, followed closely by abchampions and the clique
percolation method. However, one might object that because we cannot
assume that our ground-truth set of complexes is complete (due to the
possibility of undiscovered complexes), the NMI measure is imperfect.
\begin{table}[]

  \centering
  \begin{tabular}{|l|l l l|}
    \hline
    \textbf{Algorithm}             		&   \textbf{NMI} & $|C|$ & $Avg$ 
    \\ \hline    \hline
    % CommunityFind .. 3 0.6 1.0 2 .6
    \alg{}                         		&        0.550021 & 119 & 0.861
    \\\hline abchampions \cite{mishral-2007}  	&        0.529883 & 63 & 0.485
    \\\hline Clique percolation \cite{SCP} 	&        0.522478 & 114 & 0.744
    \\\hline MCL   \cite{vandongen-2000:2}   	&        0.414983 & 298 & 1.000
    \\\hline Blondel                	  	&        0.328344 & 212 & 1.000
    \\\hline Iterative Scan \cite{baumes-2005}	&        0.301171 & 230 & 6.975
    \\\hline COPRA    \cite{gregory-copra}	&        0.299514 & 513 & 1.101
    \\\hline LFM    \cite{lancichinetti-2009}  	&        0.270601 & 58 & 0.646
    \\ \hline   
  \end{tabular}
  \caption{NMI of various algorithms on PPI data, along with number of communities $|C|$ and average comms. per node. }
  \label{ppi-nmi}
\end{table}
% MCL has found the most communities and abchampions one of the least
% -- does this partly explain their relative F-score performance. Can
% we say anything about this in the paper?
\begin{figure}[h]
  \centering \subfigure[\alg{}]{
    \includegraphics[scale=.20]{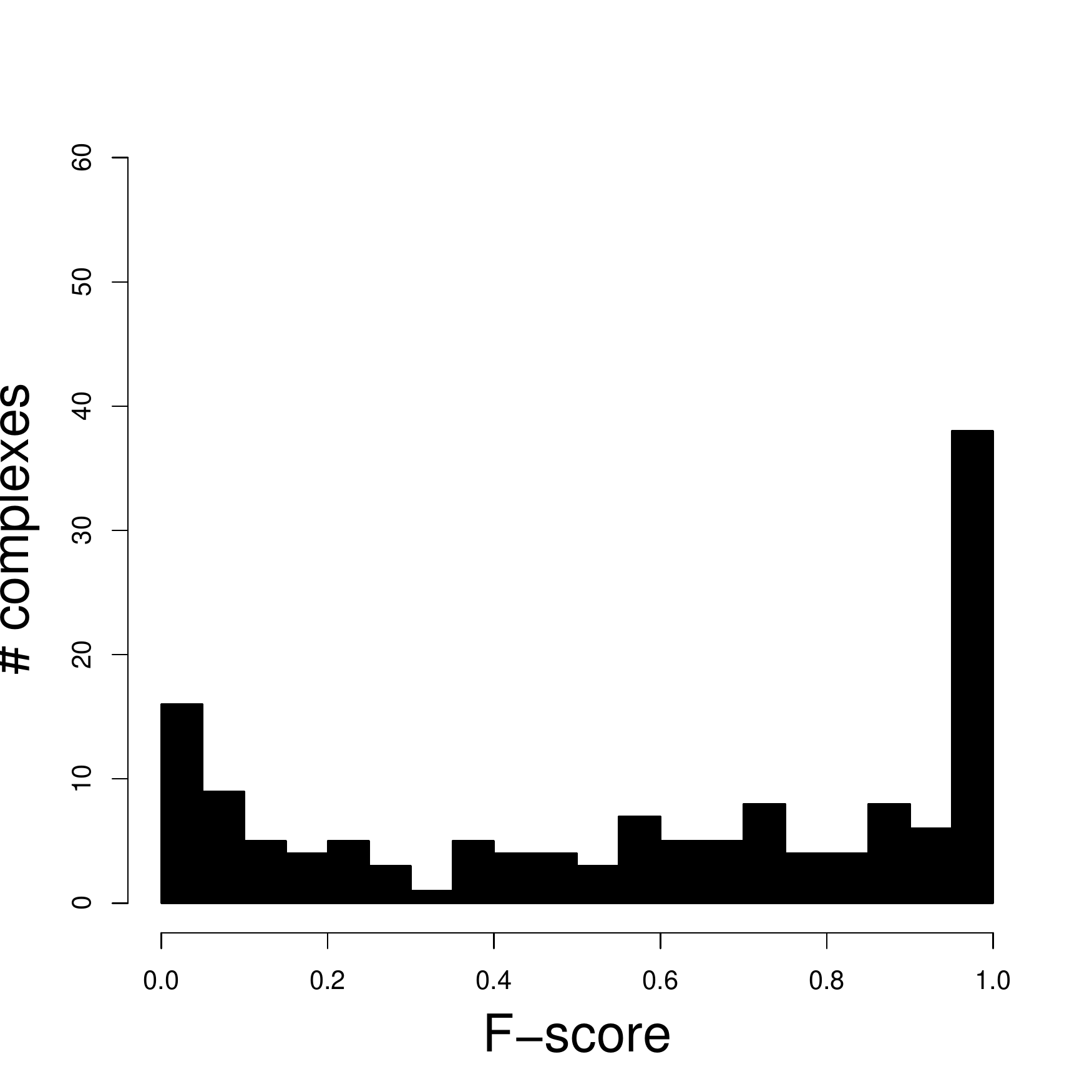}
    \label{f_score_GCE}
  } \subfigure[abchampions]{
    \includegraphics[scale=.20]{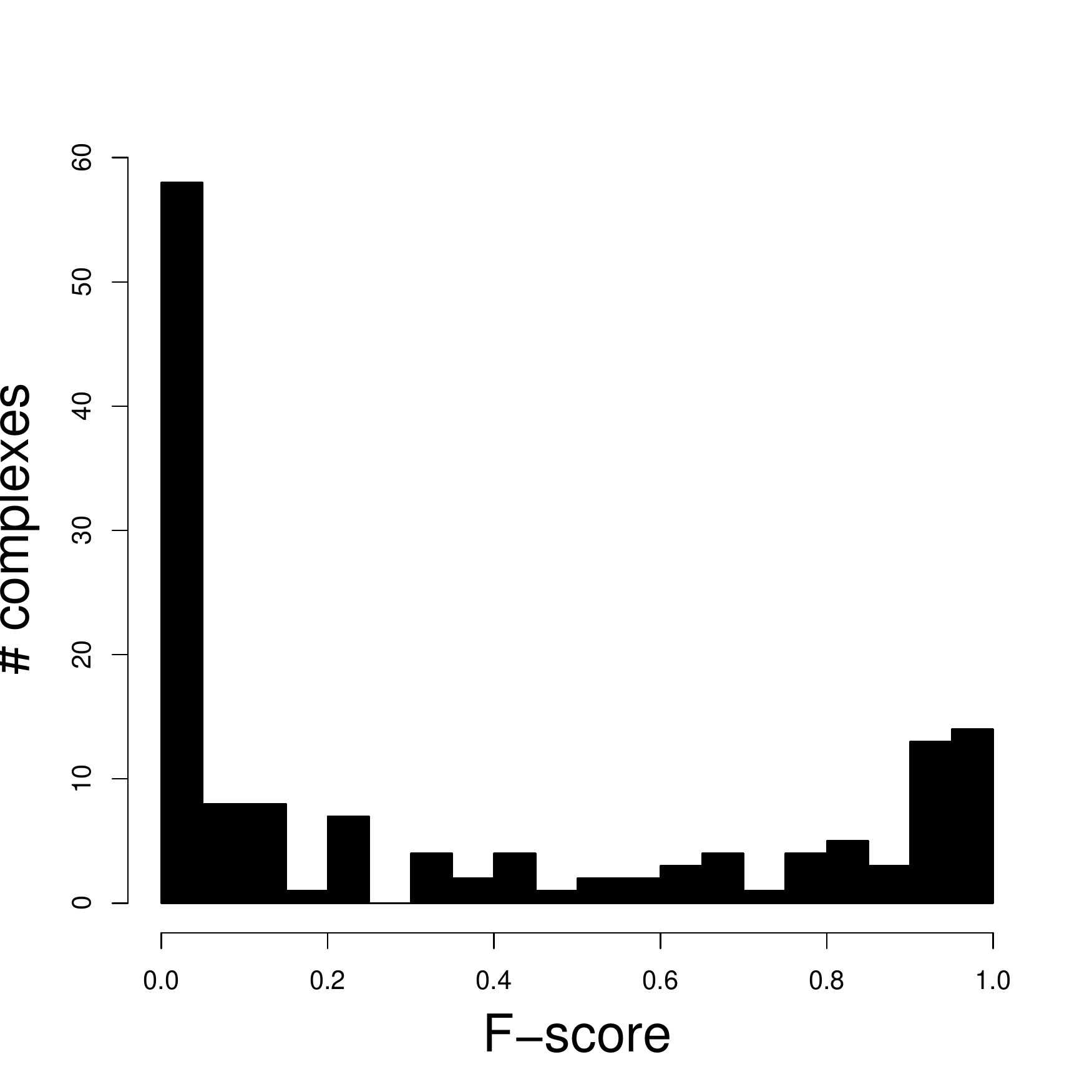}
    \label{f_score_abchampions}
  } \subfigure[SCP (k=3) ]{
    \includegraphics[scale=.20]{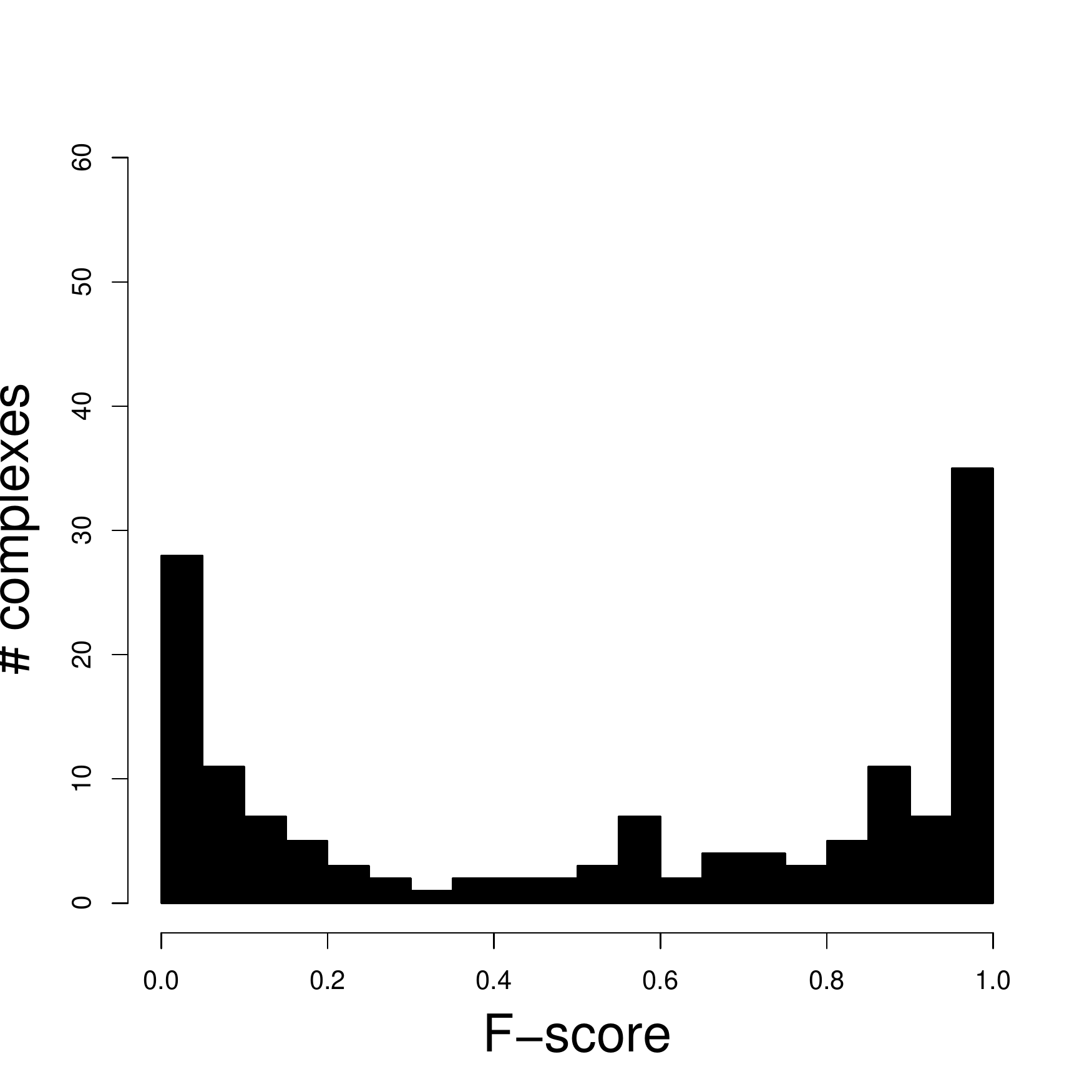}
    \label{f_score_scp}
  } \subfigure[mcl]{
    \includegraphics[scale=.20]{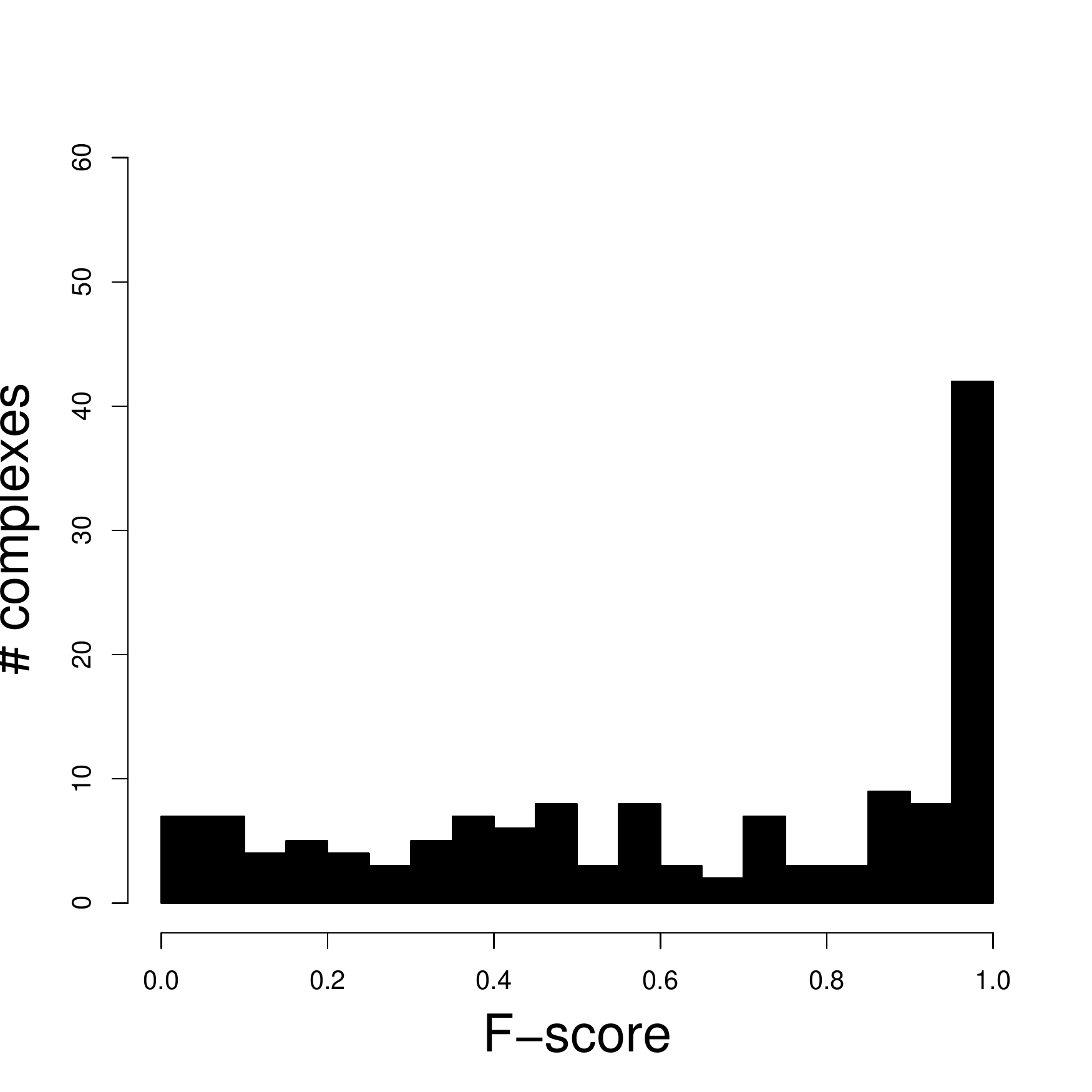}
    \label{f_score_mcl}
  }
  \caption{F-scores for each known complex in the PPI data}
  \label{PPI-Fscores}
  \vspace{-2mm}
\end{figure}

For this reason, in \cref{PPI-Fscores} we look at these PPI results in
more detail using the \Fscore, a summary of the classification metrics
recall and precision. To evaluate an CAA performance, we performed the
following procedure. For each known complex in the ground truth, we
selected the found community with the highest \Fscore. An \Fscore of
1.0 indicates a perfectly recovered protein complex. We plotted all
best \Fscores in a histogram.

The histograms show that of all the overlapping CAAs, \alg{} has the
most perfectly-recovered communities and the fewest poor matches. MCL,
a non-overlapping CAA, returned slightly better results.  Due to space
constraints, we display the \Fscore histograms of only the highest
performing algorithms.

We also looked at the second-best \Fscore to verify that the
algorithms are finding each complex once, and only once; this is
indeed the case.

\xhdr{Facebook}

As an application to social network analysis, we consider the network
of Facebook friendships \textit{between} CalTech students.
\citet{traud-2009} introduced this network data set. Their analysis
indicates that dorm assignment is particularly important to the social
life at CalTech, playing a crucial role in the formation of
communities. They argue that this data
offers a good empirical benchmark for community assignment algorithms,
using the dorm partition as an approximate ground truth. The data contains 16656 edges and 769 nodes, and
includes information on the dorm assignments of
75\% of the nodes. 

We run a number of CAAs on the CalTech Facebook network and
calculate the NMI of the communities they return with the dorm
partition. For this calculation, we leave out nodes whose dorm
assignments are unknown. The results are displayed in \cref{facebook-nmi}.

We also ran this benchmark on the
other four, much larger collegiate Facebook networks introduced by
\citeauthor{traud-2009}, and found that no algorithm
returned an NMI value greater than 0.01. A number of factors may have led
to this surprisingly low value: \citeauthor{traud-2009} indicate that
dorms play a less encompassing role in the social life of students at
these four larger universities. Some of these networks also included
one dorm attribute value with an especially large number of students;
we speculate that this value stands for ``off campus,'' and note that
the CAAs are unlikely to (nor should they) find that the thousands of off campus students all belong
together in one community, as they are listed in the ground truth. Finally, the value of the dorm assignment
was unknown for a larger proportion of nodes in these larger
data sets.

\begin{table}[htbp]
  \centering
  \begin{tabular}{|l|l|r|}
    \hline \textbf{Algorithm}          &   NMI  & time(s)
    \\\hline \alg{}                    & 0.338  & <1 
    % \\\hline \alg{} k4a1.4 & 0.275 & <1
    \\\hline Blondel                      & 0.288  & <1  
    \\\hline COPRA                        & 0.285  & <1  
    \\\hline Clique percolation           & 0.002  & 15.15 
    % \\ percolation &
    \\\hline Iterative Scan               & 0.244  & 782 
    \\\hline LFM                          & 0.007  & 23    
    \\\hline abchampions                  & 0.000  & 30
    \\\hline MCL                          & 0.159  & 4.4s
    \\ \hline   
  \end{tabular}
  \caption{NMI of the partition created by dorm assignments and the
    communites found by various CAAs when run on CalTech's Facebook
    friendship network.}
  % \vspace{-2mm}
  \label{facebook-nmi}
\end{table}

\section{Conclusion}

We have introduced an algorithm, Greedy Clique Expansion, which
combines the graph structure based approach of clique-finding methods
with the greedy expansion strategy found in other algorithms.
We demonstrate that \alg{} can
accurately recover communities on synthetic networks in which every node belongs to
four communities.  We found that no other algorithm
performed nearly as well on synthetic graphs in which every node belong to two or more communities. To determine whether these good results are robust, we
performed further comparative benchmarks on a range of LFR graphs with
non-overlapping communities, and found that \alg{} performed
competitively.  To complete our evaluation, we used \alg{} to recover
biological ground truth communities from a reference protein-protein
interaction network, and to infer non-network attributes from a social
graph. Compared to other overlapping CAAs, \alg{} gave the best
results.

\xhdr{Further Work} 
The community structure found in some networks may
exist at multiple scales due to hierarchical organization of the
system represented by the network. Ideally, a CAA would detect
community structure at all scales.  We are currently working with a
modified version of \alg{} that expands all seeds in parallel, merging
them as they become near-duplicates. By expanding all seeds to
encompass the entire graph, and merging them along the way, the algorithm
produces a dendrogram similar in some respects to the dendrograms produced by classic
agglomerative algorithms such as that of \citet{girvan-2002}, but which
allows overlap, and in which leaves represent seeds rather than nodes. Communities could
be extracted from this dendrogram by performing an analysis of
stability. This variant of
\alg{} might have the advantage of not only detecting hierachy, but also of
decreased sensitivity to the $\alpha$ parameter, because communities
are selected based on stability rather than on the first local maximum
of fitness.

We considered only a simple, greedy
expansion heuristic. Future work should investigate using more
sophisticated local heuristics that cleverly explore the most promising
sections of the search space. Furthermore, due to the local nature of
\alg{}, a parallel implementation should be straigtforward to
implement and would increase its scalability.

Finally, we would like better benchmarking abilities. On the one
hand, we need a utility that uses synthetic graphs to
systematically explore under what topological conditions the
performance of a CAA breaks down. On the other hand, we need better
empirical networks with ground truth communities to ensure that CAAs
do not merely perform well on synthetic data.

\section{Acknowledgments}
This work is supported by Science Foundation Ireland under grant
08/SRC/I1407: Clique: Graph and Network Analysis Cluster.

\end{document}